\documentclass[11pt,letterpaper]{article}
\usepackage{graphicx, epsfig} %include figure files
\usepackage{amsmath,amssymb,amsfonts,dsfont,mathrsfs,amsthm,mathtools}
\usepackage{bm} %include bold math: \bm{} creates bold letters in math mode
\usepackage[utf8]{inputenc}
\usepackage{cancel}
\usepackage[normalem]{ulem}
\usepackage{soul}
\usepackage{jheppub}
\usepackage{color}
\usepackage[usenames]{xcolor}
\usepackage{hyperref}
\usepackage{siunitx}
\hypersetup{linktocpage,colorlinks=true,urlcolor=blue,linkcolor=magenta,citecolor=blue}
\definecolor{blue-violet}{rgb}{0.54, 0.17, 0.89}
\definecolor{PineGreen}{cmyk}{0.92, 0, 0.59, 0.25}
\definecolor{Gray}{cmyk}{0, 0, 0, 0.50}

\newcommand{\diff}[1]{\text{d}#1}

\newcommand{\Lie}{\mathcal{L}}
\newcommand{\Lag}{\mathscr{L}}

\begin{document}

\title{Noether-Wald charges in six-dimensional Critical Gravity}

\author[a]{Giorgos Anastasiou,}
\author[b]{Ignacio J. Araya,} 
\author[b]{Crist\'obal Corral}
\author[c]{and Rodrigo Olea}

\emailAdd{georgios.anastasiou@pucv.cl}
\emailAdd{ignaraya@unap.cl}
\emailAdd{crcorral@unap.cl}
\emailAdd{rodrigo.olea@unab.cl}
\affiliation[a]{Instituto de F\'{\i}sica, Pontificia Universidad Cat\'{o}lica de Valpara\'{\i}so, Casilla 4059, Valpara\'{\i}so, Chile}
\affiliation[b]{Instituto de Ciencias Exactas y Naturales, Facultad de Ciencias, Universidad Arturo Prat, Avenida Arturo Prat Chac\'on 2120, 1110939, Iquique, Chile}
\affiliation[c]{Departamento de Ciencias F\'{\i}sicas, Universidad Andres Bello, Sazi\'{e} 2212, Piso 7, Santiago, Chile}

%\author{Giorgos Anastasiou}
%\affiliation{Instituto de F\'isica, Pontificia Universidad Cat\'olica de Valpara\'iso,\\ Casilla 4059, Valpara\'iso, Chile.}
%\author{Ignacio J. Araya}
%\author{Crist\'obal Corral}
%\affiliation{Instituto de Ciencias Exactas y Naturales, Facultad de Ciencias, Universidad Arturo Prat, Avenida Arturo Prat Chac\'on 2120, 1110939, Iquique, Chile}
%\author{Rodrigo Olea}
%\affiliation{Departamento de Ciencias F\'isicas, Universidad Andres Bello,\\Sazi\'e 2212, Piso 7, Santiago, Chile}

\abstract{It has been recently shown that there is a particular combination of conformal invariants in six dimensions which accepts a generic Einstein space as a solution. The Lagrangian of this Conformal Gravity theory---originally found by Lu, Pang and Pope (LPP)---can be conveniently rewritten in terms of products and covariant derivatives of the Weyl tensor. This allows one to derive the corresponding Noether prepotential and Noether-Wald charges in a compact form. Based on this expression, we calculate the Noether-Wald charges of six-dimensional Critical Gravity at the bicritical point, which is defined by the difference of the actions for Einstein-AdS gravity and the LPP Conformal Gravity. When considering Einstein manifolds, we show the vanishing of the Noether prepotential of Critical Gravity explicitly, which implies the triviality of the Noether-Wald charges. This result shows the equivalence between Einstein-AdS gravity and Conformal Gravity within its Einstein sector not only at the level of the action but also at the level of the charges.}

\maketitle

\section{Introduction}

The notion of criticality refers to the coalescence of the linearized equations of motion (EOM) such that propagating massive modes turn into massless.
In particular, Critical Gravity (CrG) in 4D is a higher-curvature theory introduced by Lu and Pope in Ref.~\cite{Lu:2011zk}, considered a probe of quantum gravity free from the usual pathologies of higher-derivative theories. As a matter of fact, the model is devoid of ghost-like modes, an appealing physical feature. The particle content of CrG is characterized by the presence of a massless graviton and the emergence of a new logarithmic mode, which can be eliminated by imposing standard asymptotically anti-de Sitter (AAdS) boundary conditions.

Critical Gravity represents a unique point in the parametric space of Einstein gravity augmented by arbitrary quadratic couplings in the curvature \cite{Lu:2011ks,Lu:2011zk}. As a consequence of this particular feature, it was shown that the mass of the Schwarzschild-AdS black hole is zero.

Later, the above result was extended to a generic Einstein space, showing that all conserved quantities are identically vanishing~\cite{Anastasiou:2017rjf}. This was understood in Ref.~\cite{Anastasiou:2016jix} as a consequence of the fact that four-dimensional renormalized Einstein-AdS gravity can be embedded into Conformal Gravity (CG) such that, for Einstein solutions, the actions of both theories are the same. In principle, the action of CG can be decomposed into Einstein and non-Einstein parts, whose dynamics is encoded in the AdS curvature
\begin{equation}\label{F}
    \mathcal{F}^{\mu\nu}_{\lambda\rho} = R^{\mu\nu}_{\lambda\rho} + \tfrac{1}{\ell^2}\delta^{\mu\nu}_{\lambda\rho} \,,
\end{equation} and the traceless Ricci tensor, respectively. The latter vanishes for Einstein spacetimes, recovering the renormalized Einstein-AdS action. Since the CrG action is constructed as the difference between the actions for Einstein-AdS and CG \cite{Lu:2011ks}, it depends entirely on the non-Einstein modes, and therefore, it yields zero upon evaluating on Einstein solutions. We recall the fact that, in the Euclidean sector, the action corresponds to the free-energy functional and thus depends on the canonical conjugates given by the black hole charges and corresponding chemical potentials. The triviality of the action, therefore, is a strong evidence of the vanishing of said conserved charges.

From the point of view of the Wald procedure~\cite{Wald:1993nt,Iyer:1994ys,Wald:1999wa}, the conserved quantities can be directly calculated following standard Noether identities. In the case of four-dimensional CrG, the analysis done in Ref.~\cite{Anastasiou:2016jix} confirmed that the only non-vanishing contribution to the Noether prepotential is coming from terms which are quadratic in the Bach tensor. This automatically implies the vanishing of the Noether charges for Einstein spacetimes, what is in agreement with the triviality of the action in this sector. In other words, the vanishing of the free energy for Einstein spacetimes turns them into an infinitely degenerate vacuum, such that this entire class becomes a vacuum modular space of the theory~\cite{Anastasiou:2017rjf}. By vacuum, it is meant the vanishing of the free energy in the Euclidean section, or alternatively, the space of configurations that have vanishing conserved charges. This is a different notion of vacuum than referring to the maximally-symmetric configuration, which still uniquely corresponds to pure AdS spacetime.

In order to further understand the role of conformal symmetry in higher-dimensional gravity, in Ref.~\cite{Anastasiou:2020mik}, novel properties of 6D Conformal Gravity were unveiled. Indeed, a particular version of CG---discovered by Lu, Pang and Pope (LPP)~\cite{Lu:2011ks}---was recently shown to contain a generic Einstein sector \cite{Anastasiou:2020mik}.
From a holographic standpoint, the action for LPP gravity reduces to the renormalized version of the corresponding Einstein functional, once generalized Neumann boundary conditions on the asymptotic metric are imposed.

In the same reference, Critical Gravity at the bicritical point in six dimensions was introduced as the difference between Einstein-AdS gravity and the newly-found CG theory that admitted Schwarzschild-AdS black holes. 

Subsequent papers contributed to the understanding of Critical Gravity from a different perspective. Taking into account the renormalization of the Einstein part of the action leads to the addition of the corresponding Euler term, which is topological in any even bulk dimension. Both terms form the renormalized Einstein-AdS gravity action. This action, in turn, was written as a polynomial of the AdS curvature in Ref.~\cite{Miskovic:2014zja}. On the other hand, by considering a compact expression of the  LPP Conformal Gravity on top of the renormalized Einstein gravity, one is able to decompose the six-dimensional Critical Gravity action in terms of the Einstein and the non-Einstein parts (proportional to the traceless Ricci tensor). Thus, the triviality of the CrG action for Einstein solutions was also made manifest in 6D.

In this work, we follow the generalized Noether-Wald procedure~\cite{Wald:1993nt,Iyer:1994ys,Wald:1999wa} to construct the Noether prepotential and compute conserved charges for AAdS solutions to 6D CG and CrG. We provide a compact form for the charges in 6D Conformal Gravity (previously studied  in Ref.~\cite{Lu:2013hx}) that turns Einstein spaces trivial within the framework of 6D Critical Gravity, as the Noether charges are identically zero.

The article is organized as follows: In Sec.~\ref{sec:CGandCrG}, we review Conformal and Critical Gravity in six dimensions; first in its seminal form (see Ref.~\cite{Lu:2011ks}) and later in their recently proposed, more compact form, in a more convenient basis~\cite{Anastasiou:2020mik}. In Sec.~\ref{sec:NW}, we present the basics of the Noether--Wald formalism and we compute the Noether prepotential for both aforementioned theories. Then, the triviality of the Noether--Wald charges of Critical Gravity in the Einstein sector is explicitly worked out. Discussion and further remarks are given in Sec.~\ref{sec:Discussion}. Additionally, we include two Appendices to render the manuscript self contained: the first one to define projections of the Riemann tensor onto different objects that  greatly simplifies the calculations and the second one for recalling properties of Einstein spacetimes.     

\section{Compact forms for the Conformal Gravity and Critical Gravity actions in six dimensions\label{sec:CGandCrG}}

\subsection{Conformal Gravity in six dimensions}

Conformal Gravity in four dimensions is constructed out of the unique conformal invariant allowed: the squared Weyl tensor. In six dimensions, however, there are three independent conformal invariants that can be used to construct the theory \cite{Bonora:1985cq,Deser:1993yx,Erdmenger:1997gy}. A linear combination of them with arbitrary coefficients does not guarantee the existence of Einstein spaces as solution to the field equations. Nevertheless, there exists a particular choice of the parameters that allows one to circumvent the previous obstruction. The conformally invariant theory possessing an Einstein-AdS sector was first obtained in Ref.~\cite{Lu:2011ks} and it is described by the action
\begin{align} \label{LPPCG}
    I_{\rm CG} = \alpha\int_{\mathcal{M}}\diff{^6x}\sqrt{-g}\;\left[\Lag_{\rm CG}^{(1)} + \frac{1}{4}\Lag_{\rm CG}^{(2)} - \frac{1}{12}\Lag_{\rm CG}^{(3)} \right] \,,
\end{align}
where
\begin{align}
    \Lag_{\rm CG}^{(1)} &= W_{\alpha\beta\mu\nu}W^{\alpha\rho\lambda\nu}W_{\rho}{}^{\beta\mu}{}_{\lambda} \,, \\
    \Lag_{\rm CG}^{(2)} &= W_{\mu\nu\alpha\beta}W^{\alpha\beta\rho\lambda}W_{\rho\lambda}{}^{\mu\nu} \,, \\
    \Lag_{\rm CG}^{(3)} &= W_{\mu\rho\sigma\lambda}\left(\delta^\mu_\nu\Box + 4R^\mu_\nu - \frac{6}{5}R\delta^\mu_\nu \right)W^{\nu\rho\sigma\lambda} + \nabla_\mu V^\mu \,, \\
    \notag
    V_\mu &= 4 R_{\mu}{}^{\lambda\rho\sigma}\nabla^\nu R_{\nu\lambda\rho\sigma} + 3R^{\nu\lambda\rho\sigma}\nabla_\mu R_{\nu\lambda\rho\sigma} - 5R^{\nu\lambda}\nabla_\mu R_{\nu\lambda} \\
    &\quad + \frac{1}{2}R\nabla_\mu R - R^\nu_\mu\nabla_\nu R + 2R^{\nu\lambda}\nabla_\nu R_{\lambda\mu} \,.
\end{align}
Here, $W^{\mu\nu}_{\lambda\rho}$ is the Weyl tensor (see Appendix~\ref{sec:irred} for conventions) and $\Lag_{\rm CG}^{(i)}$ with $i=1,2,3$ are the three conformal invariants in six dimensions built upon inequivalent contractions of the Weyl tensor and derivatives thereof. Later on, the same theory was rewritten in a more convenient and compact basis from a conformal viewpoint \cite{Anastasiou:2020mik}. In particular, the action 
\begin{align}\label{ICG}
   I_{\rm CG} &=  \alpha\int_{\mathcal{M}}\diff{^6x}\sqrt{-g}\;\left(\Lag_{\rm CG}^{(\rm bulk)} + \Lag_{\rm CG}^{(\rm bdy)}\right) \,,
\end{align}
was separated into bulk and boundary pieces given by
\begin{subequations}\label{LagCG}
\begin{align}
    \Lag_{\rm CG}^{(\rm bulk)} &= \frac{1}{4!}\delta_{\mu_1\ldots\mu_6}^{\nu_1\ldots\nu_6}W^{\mu_1\mu_2}_{\nu_1\nu_2}W^{\mu_3\mu_4}_{\nu_3\nu_4}W^{\mu_5\mu_6}_{\nu_5\nu_6} + \frac{1}{2}\delta_{\mu_1\ldots\mu_5}^{\nu_1\ldots\nu_5}W^{\mu_1\mu_2}_{\nu_1\nu_2}W^{\mu_3\mu_4}_{\nu_3\nu_4}S^{\mu_5}_{\nu_5} + 8C^{\mu\nu\lambda}C_{\mu\nu\lambda} \,, \\
    \Lag_{\rm CG}^{(\rm bdy)} &=  \nabla_\mu\left(8 W^{\mu\kappa\lambda\nu}C_{\kappa\lambda\nu} - W^{\kappa\lambda}_{\nu\sigma}\nabla^\mu W^{\nu\sigma}_{\kappa\lambda}\right) \,,
\end{align}
\end{subequations}
respectively\footnote{Here, $\delta_{\mu_1\ldots\mu_p}^{\nu_1\ldots\nu_p} = p!\,\delta_{[\mu_1}^{\nu_1}\dots\delta_{\mu_p]}^{\nu_p}$ is the generalized Kronecker delta whose trace satisfies the property
\begin{align}\label{tracedelta}
    \delta_{\mu_1\ldots\mu_p}^{\nu_1\ldots\nu_p}\delta_{\nu_1}^{\mu_1}\dots\delta_{\nu_k}^{\mu_k} = \frac{\left(D-p+k \right)!}{\left(D-p \right)!}\delta_{\mu_{k+1}\ldots\mu_p}^{\nu_{k+1}\ldots\nu_p}.
\end{align}
}. Here
\begin{equation}
C_{\mu\nu\lambda} =\nabla_{\lambda}S_{\mu\nu}- \nabla_{\nu}S_{\mu\lambda} \,,
\label{Cotton}
\end{equation}
and
\begin{equation}
S_{\nu}^{\mu}=\frac{1}{D-2}\left[ R_{\nu}^{\mu} - \frac{1}{2(D-1)}R\delta_{\nu}^{\mu}\right] \,,
\label{Schouten}
\end{equation}
are the Cotton and Schouten tensors, respectively. This form of the LPP Conformal Gravity is convenient for the study of the Einstein sector of the theory. Indeed, when the action \eqref{ICG} is evaluated in Einstein spacetimes, it reduces to the renormalized Einstein-AdS action up to a proportionality constant \cite{Anastasiou:2020mik}. This property will be crucial for the determination of the Noether-Wald charges of the theory as well as its relation to six-dimensional Critical Gravity.

\subsection{Critical Gravity in six dimensions}

The presence of special points in the parametric space of higher-curvature theories where new qualitative features arise has been reported in various examples in the literature, such as the chiral point in Topologically Massive gravity~\cite{Li:2008dq,Deser:1981wh,Deser:1982vy}, Quasitopological gravity~\cite{Oliva:2010eb,Cisterna:2017umf}, Einstenian Cubic gravity~\cite{Bueno:2016xff,Bueno:2016ypa}, Born-Infeld and Chern-Simons gravity~\cite{Banados:1993ur}, to mention a few. Critical gravity, as introduced by Lu and Pope in Ref.~\cite{Lu:2011zk}, extends the notion of criticality in four dimensions. More specifically, there is a unique combination of the Einstein-Hilbert action enhanced by a negative cosmological constant and quadratic curvature invariants that leads to the vanishing of ghost modes. The corresponding action reads
\begin{equation}
I_{\rm Crit}= \frac{1}{16 \pi G_{N}}\int_{\mathcal{M}_{4}} \diff{^4x} \left[R - 2 \Lambda + \frac{3}{2\Lambda} \left(R_{\mu \nu} R^{\mu \nu} - \frac{1}{3} R^2 \right) \right] \,.
\end{equation}
After some algebraic manipulation, the latter can be rewritten as Einstein-Weyl gravity up to the Gauss-Bonnet term which, in four dimensions, does not modify the bulk dynamics of the theory\footnote{Nevertheless, the topological term comes with a coupling constant such that, when combined with the Einstein-Hilbert action, it forms the four-dimensional renormalized Einstein-AdS action~\cite{Miskovic:2009bm}.}. Interestingly enough, this theory belongs to a generic class of quadratic curvature gravities in arbitrary dimensions which have a critical point in a unique vacuum~\cite{Deser:2011xc}. Nevertheless, 4D Critical Gravity is the only one among this class of theories admitting Einstein spacetimes as solutions.

The concept of criticality was later generalized to theories that involve non-derivative cubic curvature invariants in Ref.~\cite{Sisman:2011gz}. Even though these theories have desirable unitarity properties, it is not clear whether the solution space contains an Einstein sector, similar to the one found in four dimensions. In six dimensions, a generalization of the 4D Critical Gravity was introduced by LPP in Ref.~\cite{Lu:2011ks}. Namely, a theory with a unique vacuum that admits an Einstein sector and has massless modes in addition to the standard graviton. The LPP CrG action was given as
\begin{equation} \label{6DCrit}
I_{\rm Crit} = \frac{1}{16 \pi G_{N}}\int_{\mathcal{M}_{6}} \diff{^6x} \left[R - 2 \Lambda - \frac{\ell^4}{72} \mathcal{E}_{6}+ \frac{\ell^4}{24} \left(4 \Lag_{\rm CG}^{(1)} + \Lag_{\rm CG}^{(2)}- \frac{1}{3} \Lag_{\rm CG}^{(3)} \right) \right] \,,
\end{equation}
where
\begin{equation}
\mathcal{E}_{6} = \frac{1}{2^3} \delta_{\mu_{1} \ldots \mu_{6}}^{\nu_{1} \ldots \nu_{6}} R_{\nu_{1} \nu_{2}}^{\mu_{1} \mu_{2}} R_{\nu_{3} \nu_{4}}^{\mu_{3} \mu_{4}} R_{\nu_{5} \nu_{6}}^{\mu_{5} \mu_{6}} \,,
\end{equation}
is the topological Euler density in 6D. Similar to its 4D counterpart, Eq.~\eqref{6DCrit} becomes trivial for Schwarzschild-AdS and Kerr-AdS spacetimes, as a consequence of the equivalence between Einstein-AdS and the LPP CG action in 6D~\cite{Anastasiou:2020mik}.

As noticed in Ref.~\cite{Anastasiou:2018mfk}, considering that $\Lambda=-\frac{10}{\ell^2}$, the Einstein-Hilbert term of Eq.~\eqref{6DCrit}, together with the topological density, can be rewritten as
\begin{equation}
    \int_{\mathcal{M}_{6}} \diff{^6x} \left(R - 2 \Lambda - \frac{\ell^4}{72} \mathcal{E}_{6}\right)=\int_{\mathcal{M}_{6}}\diff{^6x}\sqrt{-g}\ell ^4P_{6}\left(\mathcal{F}\right) \,,
\end{equation}
where 
\begin{gather}
    P_{6}  \left(\mathcal{F}\right)=\frac{1}{2\left(4!\right) \ell^2} \delta_{\mu_{1} \ldots \mu_{4}}^{\nu_{1}\ldots \nu_{4}} \mathcal{F}_{\nu_{1} \nu_{2}}^{\mu_{1} \mu_{2}} \mathcal{F}_{\nu_{3} \nu_{4}}^{\mu_{3} \mu_{4}}
    -\frac{1}{\left(4!\right)^2} \delta_{\mu_{1} \ldots \mu_{6}}^{\nu_{1} \ldots \nu_{6}} \mathcal{F}_{\nu_{1} \nu_{2}}^{\mu_{1} \mu_{2}} \mathcal{F}_{\nu_{3} \nu_{4}}^{\mu_{3} \mu_{4}} \mathcal{F}_{\nu_{5} \nu_{6}}^{\mu_{5} \mu_{6}} \,,
\end{gather}
is a polynomial in the AdS curvature defined in Eq.~\eqref{F}. This combination gives the topologically renormalized Einstein-AdS gravity action~\cite{Miskovic:2014zja} for AAdS manifolds with conformally-flat boundaries~\cite{Anastasiou:2020zwc}. In order to obtain the correct renormalized action for arbitrary AAdS spacetimes, the latter action has to be supplemented by a boundary term. Therefore,
\begin{align}\notag
    I_{\rm EH}^{\rm (ren)} &= \int_{\mathcal{M}}\diff{^6x}\sqrt{-g}\Lag_{\rm EH}^{\rm (ren)} \\
    \label{IEHren}
    &= \frac{\ell^4}{16 \pi G_{N}}\left( \int_{\mathcal{M}}\diff{^6x}\sqrt{-g} P_{6}\left(\mathcal{F}\right) + \frac{1}{24} \int_{\mathcal{M}}\diff{^6x}\sqrt{-g}\,\nabla_\mu\left(\mathcal{F}^{\kappa\lambda}_{\nu\sigma}\nabla^\mu\mathcal{F}^{\nu\sigma}_{\kappa\lambda} \right)\right) \,,
\end{align}
gives the renormalized Einstein-AdS action for arbitrary 6D AAdS manifolds.

Considering $I_{\rm EH}^{\rm (ren)}$ as defined in Eq.~\eqref{IEHren} and the definition of CG given in Eq.~\eqref{LPPCG}, the CrG action of Eq.~\eqref{6DCrit} becomes
\begin{align}\label{Icrit}
    I_{\rm Crit} &= I_{\rm EH}^{\rm (ren)} - I_{\rm CG} \,,
\end{align}
where the coupling constant of CG is chosen as $\alpha = -\tfrac{\ell^4}{384\pi G_N}$. Here, unlike in the four dimensional case, the theory is not free of pathologies. Indeed, only one out of the two massive gravitons coalesces with the massless one, leaving the remaining ghost excitation intact. Thus, the action~\eqref{Icrit} corresponds to the bicritical point in the parameter space. As noted in Ref.~\cite{Lu:2011ks}, the absence of the remaining massive mode would require the addition of a Weyl squared term with fixed coupling constant. Nevertheless, since this theory (tricritical gravity) is not a linear combination of the Einstein-AdS gravity and the CG in the corresponding dimension, and since it does not have Einstein spacetimes as solutions, it is out of the scope of our analysis.

Our interest in LPP CrG, originates from the fact that Einstein-Conformal gravity at the critical point probes the emergence of Einstein gravity from Conformal Gravity~\cite{Maldacena:2011mk,Grumiller:2013mxa,Anastasiou:2016jix,Anastasiou:2020mik}. This is understood as the triviality of Critical Gravity both at the level of the action and at the level of the conserved charges for Einstein solutions~\cite{Lu:2011zk,Anastasiou:2017rjf}. The analogous analysis in six dimensions takes places at the bicritical point. The triviality of the corresponding action \eqref{ICG} was proven in Ref.~\cite{Anastasiou:2020mik}. In what follows, we study the behaviour of the Noether-Wald charges in this theory, what allows us to generalize the emergence of Einstein from Conformal Gravity in $D=6$.

\section{Noether-Wald formalism for conserved charges in Conformal Gravity and Critical Gravity\label{sec:NW}}

In this section, we review the Noether--Wald formalism for computing conserved charges\footnote{A similar formalism was previously proposed in Ref.~\cite{Nester:1991yd} (see also~\cite{Chen:1998aw}).}. Additionally, we obtain the Noether prepotential explicitly for Conformal Gravity and Critical Gravity in six dimensions. 

A $D$-dimensional action constructed out of the metric and covariant derivatives thereof, respecting diffeomorphism invariance, can be generically written as 
\begin{align}
I\left[g_{\mu\nu} \right] = \int_{\mathcal{M}}\diff{^Dx}\sqrt{-g}\;\Lag\left[g_{\mu\nu},R^{\mu\nu}_{\lambda\rho},\nabla_{\sigma_1}R^{\mu\nu}_{\lambda\rho},\ldots,\nabla_{(\sigma_1}\ldots\nabla_{\sigma_n)}R^{\mu\nu}_{\lambda\rho} \right] \,.
\end{align}
An arbitrary stationary variation of the latter with respect to the metric yields~\cite{Wald:1993nt,Iyer:1994ys,Wald:1999wa}
\begin{align}\label{INW}
    \delta I &= \int_{\mathcal{M}}\diff{^Dx}\sqrt{-g}\;\delta g^{\mu\nu}\mathcal{E}_{\mu\nu} + \int_{\mathcal{M}}\diff{^Dx}\sqrt{-g}\;\nabla_\mu \Theta^\mu \,,
\end{align}
where 
\begin{align}\label{geneom}
    \mathcal{E}_{\mu\nu} &= E_{\mu}{}^{\lambda\rho\sigma}R_{\nu\lambda\rho\sigma} - \frac{1}{2}g_{\mu\nu}\Lag - 2 \nabla^\lambda\nabla^\rho E_{\mu\lambda\rho\nu} \,,\\
    \label{genbdy}
    \Theta^\mu &= -2 \nabla_\rho\delta g_{\nu\sigma}E^{\rho\sigma\mu\nu} + 2\delta g_{\nu\sigma}\nabla_\rho E^{\rho\sigma\mu\nu}\,, \\
    \label{genE}
    E^{\mu\nu\lambda\rho} &= \frac{\partial\Lag}{\partial R_{\mu\nu\lambda\rho}} - \nabla_{\sigma_1}\frac{\partial\Lag}{\partial\nabla_{\sigma_1}R_{\mu\nu\lambda\rho}} + \ldots + (-1)^n \nabla_{(\sigma_1}\ldots\nabla_{\sigma_n)}\frac{\partial\Lag}{\partial\nabla_{(\sigma_1}\ldots\nabla_{\sigma_n)}R_{\mu\nu\lambda\rho}}\,,
\end{align}
denote the equations of motion, i.e. $\mathcal{E}_{\mu\nu}=0$, the boundary term arising from the variation, and the functional derivative of the Lagrangian with respect to the Riemann and derivatives thereof, respectively. The latter would correspond to the field equations for the Riemann tensor if it were considered as an independent field from the metric.  From hereon, we refer to $E^{\mu\nu\lambda\rho}$ as the E-tensor and we shall assume that it has the same symmetries of the Riemann tensor as in Ref.~\cite{Padmanabhan:2011ex}. 

Invariance of the action~\eqref{INW} under the diffeomorphism generated by the flow of a vector field $\xi=\xi^\mu\partial_\mu$, implies that 
\begin{align}\label{conservationlaw}
    \nabla_\mu\left[\Theta^\mu(g,\Lie_\xi g) - \Lag \xi^\mu \right] \equiv \nabla_\mu J^\mu = - \Lie_\xi g^{\mu\nu}\mathcal{E}_{\mu\nu}\,,
\end{align}
where $\Lie_\xi$ is the Lie derivative along the vector field $\xi$ and $J^\mu$ is the Noether current, i.e. 
\begin{align}
    J^\mu = -2\nabla_\nu\left(E^{\mu\nu\lambda\rho}\nabla_\lambda\xi_\rho + 2\xi_\lambda\nabla_\rho E^{\mu\nu\lambda\rho} \right) \,.
\end{align}
From Eq.~\eqref{conservationlaw}, it is direct to see that, on-shell, $\nabla_\mu J^\mu = 0$. Therefore, the Poincaré lemma allows us to express the Noether current locally as $J^\mu = \nabla_\nu q^{\mu\nu}$, where
\begin{align}\label{prepotential}
    q^{\mu\nu} = -2\left(E^{\mu\nu\lambda\rho}\nabla_\lambda \xi_\rho + 2\xi_\lambda\nabla_\rho E^{\mu\nu\lambda\rho} \right) = -q^{\nu\mu} \,,
\end{align}
is known as the Noether prepotential. In addition, if $\xi$ is a Killing vector field, one can compute conserved charge associated to the latter through
\begin{align}\label{Noethercharge}
    Q\left[\xi\right] =  \int_\Sigma q^{\mu\nu}\diff{\Sigma_{\mu\nu}}\,, %= \int_\Sigma \epsilon_{\mu_1\ldots\mu_D}q^{\mu_1\mu_2}\diff{x^{\mu_3}}\wedge\ldots\wedge\diff{x^{\mu_D}} ,
\end{align}
where $\Sigma$ is a codimension-2 hypersurface and $\diff{\Sigma_{\mu\nu}}$ its area element. 

\subsection{Noether prepotential for Conformal Gravity}

In order to compute the Noether prepotential for six dimensional Conformal Gravity, we separate the $E$-tensor~\eqref{genE} into two pieces, namely,
\begin{align}\label{ECG}
    \Big[E_{\rm CG}\Big]^{\gamma\delta}_{\sigma\tau} &= \left[E^{{\rm (bulk)}}_{\rm CG}\right]^{\gamma\delta}_{\sigma\tau} + \left[E^{{\rm (bdy)}}_{\rm CG}\right]^{\gamma\delta}_{\sigma\tau} \,.
\end{align}
The first term on the right-hand side of Eq.~\eqref{ECG} is associated to $\Lag^{\rm(bulk)}_{\rm CG}$, while the second one corresponds to $\Lag^{\rm(bdy)}_{\rm CG}$ as defined in Eq.~\eqref{LagCG}. Explicitly, they are
\begin{align}\notag
    \left[E^{{\rm (bulk)}}_{\rm CG}\right]^{\gamma\delta}_{\sigma\tau} &= \frac{\partial\Lag^{\rm(bulk)}_{\rm CG}}{\partial R^{\sigma\tau}_{\gamma\delta}} - \nabla_\rho\left[\frac{\partial\Lag^{\rm (bulk)}_{\rm CG}}{\partial\nabla_\rho R^{\sigma\tau}_{\gamma\delta}} \right]\\ \notag
    &= \frac{\partial\Lag^{\rm(bulk)}_{\rm CG}}{\partial W^{\mu\nu}_{\lambda\rho}}\frac{\partial W^{\mu\nu}_{\lambda\rho}}{\partial R^{\sigma\tau}_{\gamma\delta}} + \frac{\partial\Lag^{\rm(bulk)}_{\rm CG}}{\partial S_{\mu\nu}}\frac{\partial S_{\mu\nu}}{\partial R^{\sigma\tau}_{\gamma\delta}} - \nabla_\rho\left[\frac{\partial\Lag^{\rm(bulk)}_{\rm CG}}{\partial C_{\mu\nu\lambda}}\frac{\partial C_{\mu\nu\lambda}}{\partial \nabla_\rho R^{\sigma\tau}_{\gamma\delta}} \right] \\
    \notag
    &= \frac{\alpha}{8}\delta_{\mu\nu\mu_3\ldots\nu_6}^{\lambda\rho\nu_3\ldots\nu_6}W^{\mu_3\mu_4}_{\nu_3\nu_4}\left(W^{\mu_5\mu_6}_{\nu_5\nu_6} + 8S^{\mu_5}_{\nu_5}\delta^{\mu_6}_{\nu_6} \right)\left(\Xi^{\mu\nu}_{\lambda\rho} \right)^{\gamma\delta}_{\sigma\tau} + \frac{\alpha}{2}\delta_{\mu_1\ldots\mu_5}^{\nu_1\ldots\nu_5}W^{\mu_1\mu_2}_{\nu_1\nu_2}W^{\mu_3\mu_4}_{\nu_3\nu_4}\left(\Delta^{\mu_5}_{\nu_5}  \right)^{\gamma\delta}_{\sigma\tau} \\ 
    &\quad - 32\alpha\nabla_\rho C^{\mu\nu\rho}\left(\Delta_{\mu\nu} \right)^{\gamma\delta}_{\sigma\tau}\,,
    \label{ECGbulk}
\end{align}
and
\begin{align}\notag
    \left[E^{{\rm (bdy)}}_{\rm CG}\right]^{\gamma\delta}_{\sigma\tau} &= \frac{\partial\Lag^{\rm(bdy)}_{\rm CG}}{\partial R^{\sigma\tau}_{\gamma\delta}} - \nabla_\rho\left[\frac{\partial\Lag^{\rm (bdy)}_{\rm CG}}{\partial\nabla_\rho R^{\sigma\tau}_{\gamma\delta}}\right] + \nabla_{(\chi}\nabla_{\xi)}\left[\frac{\partial\Lag^{\rm (bdy)}_{\rm CG}}{\partial \nabla_{(\chi}\nabla_{\xi)} R^{\sigma\tau}_{\gamma\delta}}\right] \\
    \notag
    &= \frac{\partial\Lag^{\rm(bdy)}_{\rm CG}}{\partial W^{\mu\nu}_{\lambda\rho}}\frac{\partial W^{\mu\nu}_{\lambda\rho}}{\partial R^{\sigma\tau}_{\gamma\delta}} - \nabla_\rho\left[\frac{\partial\Lag^{\rm(bdy)}_{\rm CG}}{\partial\nabla_\mu W^{\kappa\lambda}_{\nu\sigma}}\frac{\partial \nabla_\mu W^{\kappa\lambda}_{\nu\sigma}}{\partial \nabla_\rho R^{\sigma\tau}_{\gamma\delta}} + \frac{\partial\Lag^{\rm(bdy)}_{\rm CG}}{\partial C_{\mu\nu\lambda}}\frac{\partial C_{\mu\nu\lambda}}{\partial\nabla_\rho R^{\sigma\tau}_{\gamma\delta}} \right] \\
    \notag 
    &\quad+ \nabla_{(\chi}\nabla_{\xi)}\left[\frac{\partial \Lag^{\rm(bdy)}_{\rm CG}}{\partial\nabla_\mu C_{\kappa\lambda\nu}}\frac{\partial \nabla_\mu C_{\kappa\lambda\nu}}{\partial \nabla_{(\chi}\nabla_{\xi)}R^{\sigma\tau}_{\gamma\delta}} + \frac{\partial \Lag^{\rm(bdy)}_{\rm CG}}{\partial \nabla_{(\alpha}\nabla_{\beta)}W^{\mu\nu}_{\kappa\lambda}}\frac{\partial \nabla_{(\alpha}\nabla_{\beta)}W^{\mu\nu}_{\kappa\lambda}}{\partial \nabla_{(\chi}\nabla_{\xi)}R^{\sigma\tau}_{\gamma\delta}} \right] \\
    &= 8\alpha\nabla_\mu C_{\nu}{}^{\lambda\rho}\left(\Xi^{\mu\nu}_{\lambda\rho} \right)^{\gamma\delta}_{\sigma\tau} + 96\alpha\nabla_\rho C^{\mu\nu\rho}\left( \Delta_{\mu\nu}\right)^{\gamma\delta}_{\sigma\tau} + 16\alpha\nabla_{(\chi}\nabla_{\xi)}W^{\chi\mu\nu\xi}\left(\Delta_{\mu\nu}\right)^{\gamma\delta}_{\sigma\tau}\,,
    \label{ECGbound}
\end{align}
where the projectors $\Xi$ and $\Delta$ have been defined in Appendix~\ref{sec:irred} and the trace of the generalized Kronecker delta in Eq.~\eqref{tracedelta} has been used to rewrite the bulk piece of the $E$-tensor. Additionally, we notice that the contribution of the second term of $\Lag_{\rm CG}^{\rm(bdy)}$ to Eq.~\eqref{ECGbound} is identically zero. Thus, we find
\begin{align}\notag
\Big[E_{\rm CG}\Big]^{\gamma\delta}_{\sigma\tau}  &=  \frac{\alpha}{8}\delta_{\mu\nu\mu_3\ldots\nu_6}^{\lambda\rho\nu_3\ldots\nu_6}W^{\mu_3\mu_4}_{\nu_3\nu_4}\left(W^{\mu_5\mu_6}_{\nu_5\nu_6} + 8S^{\mu_5}_{\nu_5}\delta^{\mu_6}_{\nu_6} \right)\left(\Xi^{\mu\nu}_{\lambda\rho} \right)^{\gamma\delta}_{\sigma\tau} + \frac{\alpha}{2}\delta_{\mu_1\ldots\mu_5}^{\nu_1\ldots\nu_5}W^{\mu_1\mu_2}_{\nu_1\nu_2}W^{\mu_3\mu_4}_{\nu_3\nu_4}\left(\Delta^{\mu_5}_{\nu_5}  \right)^{\gamma\delta}_{\sigma\tau} \\ 
    &\quad + 8\alpha\nabla_\mu C_{\nu}{}^{\lambda\rho}\left(\Xi^{\mu\nu}_{\lambda\rho} \right)^{\gamma\delta}_{\sigma\tau} + 16\alpha\nabla_\rho C^{\mu\nu\rho}\left( \Delta_{\mu\nu}\right)^{\gamma\delta}_{\sigma\tau} \,,
    \label{ECGtot}
\end{align}
where we have used Eq.~\eqref{relCandW} to express the last term of $E^{{\rm (bdy)}}_{\rm CG}$ as the covariant divergence of the Cotton tensor.

\subsection{Noether prepotential for Critical Gravity}

Similarly to CG, in order to obtain the Noether prepotential for Critical Gravity, we first need to compute the E-tensor associated to the renormalized Einstein--Hilbert action written as in Eq.~\eqref{IEHren}. Thus, using Eq.~\eqref{genE} we get
\begin{align}\notag
    \left[E^{{\rm (ren)}}_{\rm EH}\right]^{\gamma\delta}_{\sigma\tau} &= \frac{\partial \Lag_{\rm EH}^{\rm(ren)}}{\partial R^{\sigma\tau}_{\gamma\delta}} - \nabla_\rho\left[ \frac{\partial \Lag_{\rm EH}^{\rm(ren)}}{\partial \nabla_\rho R^{\sigma\tau}_{\gamma\delta}}\right] + \nabla_{(\chi}\nabla_{\rho)}\left[\frac{\partial \Lag_{\rm EH}^{\rm(ren)}}{\partial \nabla_{(\chi}\nabla_{\rho)} R^{\sigma\tau}_{\gamma\delta}}\right]  \\
    \notag 
    &= \frac{\partial \Lag_{\rm EH}^{\rm(ren)}}{\partial F^{\kappa\lambda}_{\nu\alpha}}\frac{\partial F^{\kappa\lambda}_{\nu\alpha}}{\partial R^{\sigma\tau}_{\gamma\delta}} - \nabla_\rho\left[ \frac{\partial \Lag_{\rm EH}^{\rm(ren)}}{\partial \nabla_\mu F^{\kappa\lambda}_{\nu\alpha}}\frac{\partial \nabla_\mu F^{\kappa\lambda}_{\nu\alpha}}{\partial \nabla_\rho R^{\sigma\tau}_{\gamma\delta}}\right] + \nabla_{(\chi}\nabla_{\rho)}\left[ \frac{\partial \Lag_{\rm EH}^{\rm(ren)}}{\partial \nabla_{(\beta} \nabla_{\mu)} F^{\kappa\lambda}_{\nu\alpha}}\frac{\partial \nabla_{(\beta} \nabla_{\mu)} F^{\kappa\lambda}_{\nu\alpha}}{\partial \nabla_{(\chi}\nabla_{\rho)} R^{\sigma\tau}_{\gamma\delta}}\right]
    \\
    &= - \frac{\ell^4}{384\pi G_N}\,\frac{1}{8}\delta_{\sigma\tau\mu_3\ldots\mu_6}^{\gamma\delta\nu_3\ldots\nu_6} \mathcal{F}^{\mu_3\mu_4}_{\nu_3\nu_4}\left(\mathcal{F}^{\mu_5\mu_6}_{\nu_5\nu_6} - \frac{2}{\ell^2}\delta^{\mu_5\mu_6}_{\nu_5\nu_6} \right) \,.
\end{align}
Then, the E-tensor for the Einstein-Conformal Gravity action in 6$D$ is given by 
\begin{align}\label{Ecrit}
    \Big[E_{\rm EC}\Big]^{\gamma\delta}_{\sigma\tau} &= \left[E_{\rm EH}^{\rm (ren)} \right]^{\gamma\delta}_{\sigma\tau} -  \Big[E_{\rm CG} \Big]^{\gamma\delta}_{\sigma\tau} \,,
\end{align}
for an arbitrary choice of the coupling $\alpha$ of the CG term of Eq.~\eqref{ECGtot}.

In Einstein spacetimes, the boundary E-tensor of CG vanishes identically since it involves only covariant derivatives of the Cotton tensor. On the other hand, notice that the first term on the right-hand side of Eq.~\eqref{ECGtot} involves the projection of the generalized delta onto its completely traceless part in the first four indices, whose explicit expression is given in Eq.~\eqref{deltaXi}. Thus, evaluated on Einstein spacetimes, Eq.~\eqref{ECGtot} becomes
\begin{align}\notag
    \Big[E_{\rm CG}\Big]^{\gamma\delta}_{\sigma\tau}\bigg|_{\rm E} &= \frac{\alpha}{8}\delta_{\mu_1\ldots\mu_6}^{\nu_1\ldots\nu_6}\left(\Xi^{\mu_1\mu_2}_{\nu_1\nu_2} \right)^{\gamma\delta}_{\sigma\tau}\mathcal{F}^{\mu_3\mu_4}_{\nu_5\nu_6}\left(\mathcal{F}^{\mu_5\mu_6}_{\nu_5\nu_6} - \frac{2}{\ell^2}\delta^{\mu_5\mu_6}_{\nu_5\nu_6} \right) + \frac{\alpha}{2}\delta_{\mu_1\ldots\mu_5}^{\nu_1\ldots\nu_5}\mathcal{F}^{\mu_1\mu_2}_{\nu_1\nu_2}\mathcal{F}^{\mu_3\mu_4}_{\nu_3\nu_4}\left(\Delta^{\mu_5}_{\nu_5} \right)^{\gamma\delta}_{\sigma\tau}\,, \\
    \notag
    &= \frac{\alpha}{8}\delta_{\sigma\tau\mu_3\ldots\mu_6}^{\gamma\delta\nu_3\ldots\nu_6}\mathcal{F}^{\mu_3\mu_4}_{\nu_5\nu_6}\left(\mathcal{F}^{\mu_5\mu_6}_{\nu_5\nu_6} - \frac{2}{\ell^2}\delta^{\mu_5\mu_6}_{\nu_5\nu_6} \right) \\
    \notag
    &\quad + \frac{\alpha}{8}\delta^{\nu_1\ldots\nu_5}_{\mu_1\ldots\mu_5}\left(\frac{1}{20}\delta_{\nu_5}^{\mu_5}\delta^{\gamma\delta}_{\sigma\tau}  + \delta^{\mu_5}_{[\sigma}\delta_{\tau]}^{[\gamma}\delta^{\delta]}_{\nu_5} \right)\mathcal{F}^{\mu_1\mu_2}_{\nu_1\nu_2}\left(\mathcal{F}^{\mu_3\mu_4}_{\nu_3\nu_4} - \frac{2}{\ell^2}\delta^{\mu_3\mu_4}_{\nu_3\nu_4} \right) \\
    \notag
    &\quad -\frac{\alpha}{16}\delta_{\mu_1\ldots\mu_5}^{\nu_1\ldots\nu_5}\left(2\delta^{\mu_5}_{[\sigma}\delta_{\tau]}^{[\gamma}\delta^{\delta]}_{\nu_5} + \frac{1}{10}\delta^{\mu_5}_{\nu_5}\delta^{\gamma\delta}_{\sigma\tau} \right)\mathcal{F}^{\mu_1\mu_2}_{\nu_1\nu_2}\mathcal{F}^{\mu_3\mu_4}_{\nu_3\nu_4} \,,
    \end{align}
where the definition of $\Delta$ in Eq.~\eqref{Deltadef} has been used. After some algebra the last expression gives
\begin{equation}
  \Big[E_{\rm CG}\Big]^{\gamma\delta}_{\sigma\tau}\bigg|_{\rm E}  = \frac{\alpha}{8}\delta_{\sigma\tau\mu_3\ldots\mu_6}^{\gamma\delta\nu_3\ldots\nu_6}\mathcal{F}^{\mu_3\mu_4}_{\nu_5\nu_6}\left(\mathcal{F}^{\mu_5\mu_6}_{\nu_5\nu_6} - \frac{2}{\ell^2}\delta^{\mu_5\mu_6}_{\nu_5\nu_6} \right)  \,.
    \label{ECGCritical}
\end{equation}
 Therefore, for Einstein-Conformal Gravity on Einstein spacetimes we find 
\begin{align}
    \Big[E_{\rm EC} \Big]^{\gamma\delta}_{\sigma\tau}\bigg|_{\rm E} &= \left(-\frac{\ell^4}{384\pi G_N} - \alpha \right)\frac{1}{8}\delta_{\sigma\tau\mu_3\ldots\mu_6}^{\gamma\delta\nu_3\ldots\nu_6}\mathcal{F}^{\mu_3\mu_4}_{\nu_5\nu_6}\left(\mathcal{F}^{\mu_5\mu_6}_{\nu_5\nu_6} - \frac{2}{\ell^2}\delta^{\mu_5\mu_6}_{\nu_5\nu_6} \right)\,.
\end{align}
Notice that the latter vanishes identically at the critical point of the parameter space where the CrG theory of Eq.~\eqref{Icrit} is defined, namely, when $\alpha=-\tfrac{\ell^4}{384\pi G_N}$. Since the Noether prepotential is defined through Eq.~\eqref{prepotential}, it is clear that it becomes zero for Einstein spacetimes at the critical point and so do the corresponding Noether-Wald conserved charges.

This result indicates the triviality of the Einstein sector of Einstein-Conformal gravity at the bicritical point, which corresponds to the CrG theory. Indeed, it is a direct consequence of the coincidence between the Noether-Wald charges of Einstein-AdS gravity and LPP CG when the coupling of the latter is $\alpha=-\tfrac{\ell^4}{384\pi G_N}$. Thus, the two theories are not only equivalent at the level of the action but also at the variation of it.

\section{Discussion\label{sec:Discussion}}

In this work, we used the Noether-Wald formalism~\cite{Iyer:1994ys} to compute the prepotential $q^{\mu \nu}$ for the unique six-dimensional Conformal Gravity with an Einstein sector~\cite{Lu:2011ks}. A key ingredient in this computation, is the recasting of the LPP action in terms of the Weyl, Schouten and Cotton tensors~\cite{Anastasiou:2020mik}, which has a compact form that significantly streamlines the derivation.

The contributions to the prepotential coming from the bulk and the boundary (Eqs.~\eqref{ECGbulk} and~\eqref{ECGbound}, respectively) were found and written in a simpler form, introducing two tensors, $\Xi$ and $\Delta$, defined in Eqs.~\eqref{Xidef} and~\eqref{Deltadef}, respectively. These objects act as projection operators when contracted with the curvature tensor (see Appendix~\ref{sec:irred}), such that they give rise to the Weyl and Schouten tensors, respectively. They can be understood as operators which reduce the Riemann tensor to its totally traceless and partially traceless parts, such that the full Riemann admits an expansion in terms of these irreducible pieces.

Based on our result for Conformal Gravity, we then obtained the Noether prepotential for Critical Gravity at the bicritical point. The action of the theory is defined as the difference between the actions for renormalized Einstein-AdS gravity and CG in six dimensions, with the overall coupling $\alpha$ of CG chosen such that the $I_{EH}^{\text{(ren)}}$ action is recovered for Einstein spaces~\cite{Lu:2011ks,Anastasiou:2020mik}. Therefore, its action vanishes identically for Einstein solutions.

A general connection between the vanishing of the action for a class of solutions and the vanishing of the corresponding charges is a subject not commonly discussed in the literature. One may provide support to this idea considering gravity actions augmented by topological terms and defining NW charges for the total action. Applying this principle to the Einstein-AdS action in even bulk dimensions, say $D=2n$, the action rendered finite by the addition of the Euler term turns into a polynomial of degree $n$ in the Weyl tensor. As a consequence, the Noether charge is a polynomial of degree $n-1$. This immediately implies the vanishing of both the action and the charges for maximally symmetric spaces~\cite{Miskovic:2014zja}. Another illustrative example is given by the addition of both Gauss-Bonnet and Chern-Pontryagin invariants to the $4D$ Einstein-AdS gravity. The direct application of NW procedure leads to a charge which is identically vanishing for globally (anti-)self dual solutions, feature that is shared by the total action~\cite{Araneda:2016iiy,Ciambelli:2020qny}. This reasoning opens the possibility of zero action/zero charges for the whole class of spaces, and not only for a particular solution. As a matter of fact, that was also the case of CrG in four dimensions \cite{Anastasiou:2017rjf,Anastasiou:2017mag}.

It was then expected that, for Einstein manifolds, the prepotential in 6D Critical Gravity and thus, its asymptotic charges, were vanishing as well. We explicitly checked the triviality of Eq.~\eqref{prepotential} for the Einstein case, thus confirming this expectation.

The fact that the Noether charges are zero for Einstein solutions of six-dimensional bi-critical gravity, together with the vanishing of its action, are in complete analogy with the four-dimensional case studied in \cite{Lu:2011zk,Anastasiou:2016jix,Anastasiou:2017rjf,Lu:2011ks}. It is interesting to ponder on the implications, as from a thermodynamic point of view, the result suggests that Critical Gravity has a degenerate vacuum sector such that it accommodates the entire Einstein class, having vanishing free energy. Indeed, not all Einstein spaces are maximally symmetric. However, from the point of view of the thermodynamic potential, the maximally symmetric configuration---pure AdS---is neither preferred, nor stabler, but merely a representative of the Einstein vacuum. Although certainly interesting, an analysis of possible effects such as spontaneous transitions between Einstein configurations in the context of Critical Gravity, reminiscent of the Hawking-Page mechanism, are beyond the scope of the present paper.

\acknowledgments

The work of IJA is funded by Agencia Nacional de Investigación y Desarrollo (ANID), REC Convocatoria Nacional Subvenci\'on a Instalaci\'on en la Academia Convocatoria A\~no 2020, Folio PAI77200097. This work is partially funded by ANID through FONDECYT Grants No~3190314 (GA), No~11200025 (CC) and No~1170765 (RO).

\appendix

\section{Irreducible decomposition of the Riemann tensor and useful formulae\label{sec:irred}}

The Riemann tensor can be written in terms of its irreducible components as
\begin{align}
    R^{\mu\nu}_{\lambda\rho} &= A^{\mu\nu}_{\lambda\rho} + H^{\mu\nu}_{\lambda\rho} + W^{\mu\nu}_{\lambda\rho} \,,
\end{align}
where the trace, partially traceless and totally traceless parts are 
\begin{align}
    A^{\mu\nu}_{\lambda\rho} &= \frac{1}{D(D-1)}\delta_{\lambda\rho}^{\mu\nu} R \,, \\
    H^{\mu\nu}_{\lambda\rho} &= \frac{4}{D-2}\delta_{[\lambda}^{[\mu}H_{\rho]}^{\nu]} \,,\\
    W^{\mu\nu}_{\lambda\rho} &= R^{\mu\nu}_{\lambda\rho} - A^{\mu\nu}_{\lambda\rho} - H^{\mu\nu}_{\lambda\rho}\,,
\end{align}
respectively.
Here, $H_{\mu\nu} = R_{\mu\nu} - \tfrac{1}{D}g_{\mu\nu}R$ is the traceless Ricci tensor while $W^{\mu\nu}_{\lambda\rho}$ is the Weyl tensor.

It is useful to rewrite the Ricci scalar, Ricci tensor, and traceless Ricci tensor as projections of the Riemann tensor according to
\begin{align}
    R &= \frac{1}{2} \delta_{\sigma\tau}^{\gamma\delta} R^{\sigma\tau}_{\gamma\delta}\,, & R^{\mu}_{\nu} &= \delta^\mu_{[\tau}\delta_{\sigma]}^{[\gamma}\delta^{\delta]}_\nu R^{\sigma\tau}_{\gamma\delta}\,, &
    H^\mu_\nu &= \left(\delta^\mu_{[\tau}\delta_{\sigma]}^{[\gamma}\delta^{\delta]}_\nu  - \frac{1}{2D}\delta^\mu_\nu\delta_{\sigma\tau}^{\gamma\delta}\right) R^{\sigma\tau}_{\gamma\delta}\,,
\end{align}
respectively. The Schouten tensor, on the other hand, is defined through
\begin{align}\notag
    S^\mu_\nu &= \frac{1}{D-2}\left(R^\mu_\nu - \frac{1}{2(D-1)}\delta^\mu_\nu R \right) \\
    \label{Deltadef}
    &= -\frac{1}{D-2}\left(\delta^\mu_{[\sigma}\delta_{\tau]}^{[\gamma}\delta^{\delta]}_\nu  + \frac{1}{4(D-1)}\delta^\mu_\nu\delta_{\sigma\tau}^{\gamma\delta}\right)R^{\sigma\tau}_{\gamma\delta} \equiv \left(\Delta^\mu_\nu\right)^{\gamma\delta}_{\sigma\tau} R^{\sigma\tau}_{\gamma\delta} \,,
\end{align}
and it can be used to define the Cotton tensor, namely,
\begin{align}\notag
C_{\mu\nu\lambda} = 2\nabla_{[\lambda}S_{\nu]\mu} &= -\frac{1}{D-2} \delta_{\lambda\nu}^{\rho\alpha}\delta_\mu^\beta \left( g_{\alpha[\sigma}\delta_{\tau]}^{[\gamma}\delta^{\delta]}_{\beta}  + \frac{1}{4(D-1)}g_{\alpha\beta}\delta_{\sigma\tau}^{\gamma\delta}\right) \nabla_\rho R^{\sigma\tau}_{\gamma\delta} \\
&= \delta_{\lambda\nu}^{\rho\alpha} \left(\Delta_{\mu\alpha}\right)^{\gamma\delta}_{\sigma\tau} \nabla_\rho R^{\sigma\tau}_{\gamma\delta}\,,
\end{align}
where $\left(\Delta_{\mu\alpha} \right)^{\gamma\delta}_{\sigma\tau} = g_{\alpha\nu}\left(\Delta_{\mu}^\nu \right)^{\gamma\delta}_{\sigma\tau}$. Similarly, the Weyl tensor can be written in terms of projection of the Riemann tensor as
\begin{align}\notag
W^{\mu\nu}_{\lambda\rho} &=\frac{1}{4} \left(\delta_{\sigma\tau}^{\mu\nu}\delta_{\lambda\rho}^{\gamma\delta} - \frac{16}{D-2} \delta_{[\lambda}^{[\delta}\delta_{\rho]}^{[\mu}\delta_{[\sigma}^{\nu]}\delta_{\tau]}^{\gamma]} + \frac{2}{\left(D-1 \right)\left(D-2 \right)}\delta_{\lambda\rho}^{\mu\nu}\delta_{\sigma\tau}^{\gamma\delta} \right)  R^{\sigma\tau}_{\gamma\delta} \\
\label{Xidef}
&\equiv \left(\Xi^{\mu\nu}_{\lambda\rho}\right)^{\gamma\delta}_{\sigma\tau} R^{\sigma\tau}_{\gamma\delta} \,.
\end{align}
Thus, it is clear that the role of $\Xi$ is to project the Riemann tensor onto its completely traceless part. Indeed, from its symmetries, one can explicitly check that
\begin{align}
\left(\Xi^{\mu\nu}_{\lambda\rho}\right)^{\gamma\delta}_{\sigma\tau} = \left(\Xi^{\gamma\delta}_{\sigma\tau}\right)^{\mu\nu}_{\lambda\rho}\,.  
\end{align}
Therefore, since the Weyl tensor is traceless in any pair of indices, we have 
\begin{align}\label{Wontoitself}
    \left(\Xi^{\mu\nu}_{\lambda\rho}\right)^{\gamma\delta}_{\sigma\tau} W^{\sigma\tau}_{\gamma\delta} = W^{\mu\nu}_{\lambda\rho}\,,
\end{align}
that leads directly to the idempotency of this projector, namely,
\begin{align}
    \left(\Xi^{\mu\nu}_{\lambda\rho}\right)^{\gamma\delta}_{\sigma\tau}\left(\Xi^{\sigma\tau}_{\gamma\delta}\right)^{\alpha\beta}_{\chi\xi} = \left(\Xi^{\mu\nu}_{\lambda\rho}\right)^{\gamma\delta}_{\sigma\tau}\left(\Xi^{\alpha\beta}_{\chi\xi}\right)^{\sigma\tau}_{\gamma\delta} = \left(\Xi^{\mu\nu}_{\lambda\rho}\right)^{\alpha\beta}_{\chi\xi} = \left(\Xi^{\alpha\beta}_{\chi\xi}\right)^{\mu\nu}_{\lambda\rho}\,.
\end{align}
Moreover, one can find a relation between the projectors $\Xi$ and $\Delta$ from the relation between the Cotton and the Weyl tensor, that is
\begin{align}\label{relCandW}
  C_{\mu\nu\lambda} = -\frac{1}{D-3}\nabla_\rho W^{\rho}{}_{\mu\nu\lambda}\,,
\end{align}
giving
\begin{align}\label{DeltaXi}
\left(\Delta^\mu_\nu\right)^{\gamma\delta}_{\sigma\tau} = \frac{1}{(D-1)(D-3)}\left(\Xi^{\rho\mu}_{\rho\nu} \right)^{\gamma\delta}_{\sigma\tau}\,.    
\end{align}
Comparing Eqs.~\eqref{Deltadef} and~\eqref{DeltaXi} we find explicitly that
\begin{align}
    \left(\Xi^{\rho\mu}_{\rho\nu} \right)^{\gamma\delta}_{\sigma\tau} &= -\frac{(D-1)(D-3)}{(D-2)}\left(\delta^{\mu}_{[\sigma}\delta_{\tau]}^{[\gamma}\delta^{\delta]}_{\nu} + \frac{1}{4(D-1)}\delta_{\nu}^{\mu}\delta_{\sigma\tau}^{\gamma\delta} \right)\,.
\end{align}
Additionally, the projection onto the completely traceless part in the first four indices of the generalized delta gives
\begin{align}\label{deltaXi}
    \delta_{\mu_1\ldots\mu_p}^{\nu_1\ldots\nu_p}\left(\Xi^{\mu_1\mu_2}_{\nu_1\nu_2}\right)^{\gamma\delta}_{\sigma\tau} &= \delta_{\sigma\tau\mu_3\ldots\mu_p}^{\gamma\delta\nu_3\ldots\nu_p} + \frac{4(D-p+1)}{D-2}\delta_{\mu_2\ldots\mu_p}^{\nu_2\ldots\nu_p}\left(\delta^{\mu_2}_{[\sigma}\delta_{\tau]}^{[\gamma}\delta_{\nu_2}^{\delta]} +  \frac{1}{4(D-1)}\delta^{\mu_2}_{\nu_2}\delta_{\sigma\tau}^{\gamma\delta} \right)\,.
    %\label{deltaDelta}
    %\delta_{\mu_1\ldots\mu_p}^{\nu_1\ldots\nu_p}\left(\Delta^{\mu_1}_{\nu_1} \right)^{\gamma\delta}_{\sigma\tau} &= -\frac{1}{2(D-2)}\delta_{\mu_1\ldots\mu_p}^{\nu_1\ldots\nu_p}\left(\delta^{\mu_1}_{[\sigma}\delta_{\tau]}^{[\gamma}\delta^{\delta]}_{\nu_1} + g_{\nu_1[\sigma}\delta_{\tau]}^{[\gamma}g^{\delta]\mu_1} + \frac{1}{2(D-1)}\delta_{\nu_1}^{\mu_1}\delta_{\sigma\tau}^{\gamma\delta} \right).
\end{align}

\section{Einstein spacetimes\label{sec:Einstein-AdS}}

In this Appendix, we briefly review the definition of Einstein spacetimes, which are defined as solution to the Einstein field equations in presence of cosmological constant in $D$ dimensions,
\begin{align}
    R_{\mu\nu} - \frac{1}{2}R\,g_{\mu\nu} + \Lambda g_{\mu\nu} = 0 \,.
\end{align}
Taking the trace, it is direct to see that this equation is solved by
\begin{align}\label{Espace}
    R_{\mu\nu} = - \frac{(D-1)}{\ell^2}g_{\mu\nu} \,,
\end{align}
where $\ell$ is the AdS radius, related to the cosmological constant through
\begin{align}
    \Lambda = - \frac{(D-1)(D-2)}{2\ell^2} \,.
\end{align}
Thus, the condition~\eqref{Espace} defines an Einstein spacetime. Notice that, when Eq.~\eqref{Espace} holds, the traceless Ricci vanishes identically, namely, $H_{\mu\nu}=0$. This, in turn, implies that the Schouten tensor evaluated at Einstein-AdS spacetimes becomes
\begin{align}
    S^\mu_\nu = - \frac{1}{2\ell^2}\delta^\mu_\nu \,,
\end{align}
and the Cotton tensor is zero as a consequence of its definition $C_{\mu\nu\lambda} = 2\nabla_{[\lambda}S_{\nu]\mu}$. Additionally, the Weyl tensor for Einstein-AdS spacetimes becomes the AdS curvature, namely,
\begin{align}
    W^{\mu\nu}_{\lambda\rho} = R^{\mu\nu}_{\lambda\rho} + \frac{1}{\ell^2}\delta^{\mu\nu}_{\lambda\rho} \equiv \mathcal{F}^{\mu\nu}_{\lambda\rho} \,.
\end{align}

\bibliography{6DNW}

\providecommand{\href}[2]{#2}\begingroup\raggedright\begin{thebibliography}{10}

\bibitem{Lu:2011zk}
H.~Lu and C.~N. Pope, \emph{{Critical Gravity in Four Dimensions}},
  \href{http://dx.doi.org/10.1103/PhysRevLett.106.181302}{\emph{Phys. Rev.
  Lett.} {\bfseries 106} (2011) 181302},
  [\href{https://arxiv.org/abs/1101.1971}{{\ttfamily 1101.1971}}].

\bibitem{Lu:2011ks}
H.~Lu, Y.~Pang and C.~N. Pope, \emph{{Conformal Gravity and Extensions of
  Critical Gravity}},
  \href{http://dx.doi.org/10.1103/PhysRevD.84.064001}{\emph{Phys. Rev.}
  {\bfseries D84} (2011) 064001},
  [\href{https://arxiv.org/abs/1106.4657}{{\ttfamily 1106.4657}}].

\bibitem{Anastasiou:2017rjf}
G.~Anastasiou, R.~Olea and D.~Rivera~Betancour, \emph{{Noether-Wald energy in
  Critical Gravity}},  \href{https://arxiv.org/abs/1707.00341}{{\ttfamily
  1707.00341}}.

\bibitem{Anastasiou:2016jix}
G.~Anastasiou and R.~Olea, \emph{{From conformal to Einstein Gravity}},
  \href{http://dx.doi.org/10.1103/PhysRevD.94.086008}{\emph{Phys. Rev.}
  {\bfseries D94} (2016) 086008},
  [\href{https://arxiv.org/abs/1608.07826}{{\ttfamily 1608.07826}}].

\bibitem{Wald:1993nt}
R.~M. Wald, \emph{{Black hole entropy is the Noether charge}},
  \href{http://dx.doi.org/10.1103/PhysRevD.48.R3427}{\emph{Phys. Rev. D}
  {\bfseries 48} (1993) R3427--R3431},
  [\href{https://arxiv.org/abs/gr-qc/9307038}{{\ttfamily gr-qc/9307038}}].

\bibitem{Iyer:1994ys}
V.~Iyer and R.~M. Wald, \emph{{Some properties of Noether charge and a proposal
  for dynamical black hole entropy}},
  \href{http://dx.doi.org/10.1103/PhysRevD.50.846}{\emph{Phys. Rev. D}
  {\bfseries 50} (1994) 846--864},
  [\href{https://arxiv.org/abs/gr-qc/9403028}{{\ttfamily gr-qc/9403028}}].

\bibitem{Wald:1999wa}
R.~M. Wald and A.~Zoupas, \emph{{A General definition of 'conserved quantities'
  in general relativity and other theories of gravity}},
  \href{http://dx.doi.org/10.1103/PhysRevD.61.084027}{\emph{Phys. Rev. D}
  {\bfseries 61} (2000) 084027},
  [\href{https://arxiv.org/abs/gr-qc/9911095}{{\ttfamily gr-qc/9911095}}].

\bibitem{Anastasiou:2020mik}
G.~Anastasiou, I.~J. Araya and R.~Olea, \emph{{Einstein Gravity from Conformal
  Gravity in 6D}}, \href{http://dx.doi.org/10.1007/JHEP01(2021)134}{\emph{JHEP}
  {\bfseries 01} (2021) 134},
  [\href{https://arxiv.org/abs/2010.15146}{{\ttfamily 2010.15146}}].

\bibitem{Miskovic:2014zja}
O.~Miskovic, R.~Olea and M.~Tsoukalas, \emph{{Renormalized AdS action and
  Critical Gravity}},
  \href{http://dx.doi.org/10.1007/JHEP08(2014)108}{\emph{JHEP} {\bfseries 08}
  (2014) 108}, [\href{https://arxiv.org/abs/1404.5993}{{\ttfamily 1404.5993}}].

\bibitem{Lu:2013hx}
H.~L\"u, Y.~Pang and C.~N. Pope, \emph{{Black Holes in Six-dimensional
  Conformal Gravity}},
  \href{http://dx.doi.org/10.1103/PhysRevD.87.104013}{\emph{Phys. Rev. D}
  {\bfseries 87} (2013) 104013},
  [\href{https://arxiv.org/abs/1301.7083}{{\ttfamily 1301.7083}}].

\bibitem{Bonora:1985cq}
L.~Bonora, P.~Pasti and M.~Bregola, \emph{{WEYL COCYCLES}},
  \href{http://dx.doi.org/10.1088/0264-9381/3/4/018}{\emph{Class. Quant. Grav.}
  {\bfseries 3} (1986) 635}.

\bibitem{Deser:1993yx}
S.~Deser and A.~Schwimmer, \emph{{Geometric classification of conformal
  anomalies in arbitrary dimensions}},
  \href{http://dx.doi.org/10.1016/0370-2693(93)90934-A}{\emph{Phys. Lett. B}
  {\bfseries 309} (1993) 279--284},
  [\href{https://arxiv.org/abs/hep-th/9302047}{{\ttfamily hep-th/9302047}}].

\bibitem{Erdmenger:1997gy}
J.~Erdmenger, \emph{{Conformally covariant differential operators: Properties
  and applications}},
  \href{http://dx.doi.org/10.1088/0264-9381/14/8/008}{\emph{Class. Quant.
  Grav.} {\bfseries 14} (1997) 2061--2084},
  [\href{https://arxiv.org/abs/hep-th/9704108}{{\ttfamily hep-th/9704108}}].

\bibitem{Li:2008dq}
W.~Li, W.~Song and A.~Strominger, \emph{{Chiral Gravity in Three Dimensions}},
  \href{http://dx.doi.org/10.1088/1126-6708/2008/04/082}{\emph{JHEP} {\bfseries
  04} (2008) 082}, [\href{https://arxiv.org/abs/0801.4566}{{\ttfamily
  0801.4566}}].

\bibitem{Deser:1981wh}
S.~Deser, R.~Jackiw and S.~Templeton, \emph{{Topologically Massive Gauge
  Theories}},
  \href{http://dx.doi.org/10.1016/0003-4916(82)90164-6}{\emph{Annals Phys.}
  {\bfseries 140} (1982) 372--411}.

\bibitem{Deser:1982vy}
S.~Deser, R.~Jackiw and S.~Templeton, \emph{{Three-Dimensional Massive Gauge
  Theories}}, \href{http://dx.doi.org/10.1103/PhysRevLett.48.975}{\emph{Phys.
  Rev. Lett.} {\bfseries 48} (1982) 975--978}.

\bibitem{Oliva:2010eb}
J.~Oliva and S.~Ray, \emph{{A new cubic theory of gravity in five dimensions:
  Black hole, Birkhoff's theorem and C-function}},
  \href{http://dx.doi.org/10.1088/0264-9381/27/22/225002}{\emph{Class. Quant.
  Grav.} {\bfseries 27} (2010) 225002},
  [\href{https://arxiv.org/abs/1003.4773}{{\ttfamily 1003.4773}}].

\bibitem{Cisterna:2017umf}
A.~Cisterna, L.~Guajardo, M.~Hassaine and J.~Oliva, \emph{{Quintic
  quasi-topological gravity}},
  \href{http://dx.doi.org/10.1007/JHEP04(2017)066}{\emph{JHEP} {\bfseries 04}
  (2017) 066}, [\href{https://arxiv.org/abs/1702.04676}{{\ttfamily
  1702.04676}}].

\bibitem{Bueno:2016xff}
P.~Bueno and P.~A. Cano, \emph{{Einsteinian cubic gravity}},
  \href{http://dx.doi.org/10.1103/PhysRevD.94.104005}{\emph{Phys. Rev. D}
  {\bfseries 94} (2016) 104005},
  [\href{https://arxiv.org/abs/1607.06463}{{\ttfamily 1607.06463}}].

\bibitem{Bueno:2016ypa}
P.~Bueno, P.~A. Cano, V.~S. Min and M.~R. Visser, \emph{{Aspects of general
  higher-order gravities}},
  \href{http://dx.doi.org/10.1103/PhysRevD.95.044010}{\emph{Phys. Rev. D}
  {\bfseries 95} (2017) 044010},
  [\href{https://arxiv.org/abs/1610.08519}{{\ttfamily 1610.08519}}].

\bibitem{Banados:1993ur}
M.~Banados, C.~Teitelboim and J.~Zanelli, \emph{{Dimensionally continued black
  holes}}, \href{http://dx.doi.org/10.1103/PhysRevD.49.975}{\emph{Phys. Rev. D}
  {\bfseries 49} (1994) 975--986},
  [\href{https://arxiv.org/abs/gr-qc/9307033}{{\ttfamily gr-qc/9307033}}].

\bibitem{Miskovic:2009bm}
O.~Miskovic and R.~Olea, \emph{{Topological regularization and self-duality in
  four-dimensional anti-de Sitter gravity}},
  \href{http://dx.doi.org/10.1103/PhysRevD.79.124020}{\emph{Phys. Rev.}
  {\bfseries D79} (2009) 124020},
  [\href{https://arxiv.org/abs/0902.2082}{{\ttfamily 0902.2082}}].

\bibitem{Deser:2011xc}
S.~Deser, H.~Liu, H.~Lu, C.~N. Pope, T.~C. Sisman and B.~Tekin, \emph{{Critical
  Points of D-Dimensional Extended Gravities}},
  \href{http://dx.doi.org/10.1103/PhysRevD.83.061502}{\emph{Phys. Rev. D}
  {\bfseries 83} (2011) 061502},
  [\href{https://arxiv.org/abs/1101.4009}{{\ttfamily 1101.4009}}].

\bibitem{Sisman:2011gz}
T.~C. Sisman, I.~Gullu and B.~Tekin, \emph{{All unitary cubic curvature
  gravities in D dimensions}},
  \href{http://dx.doi.org/10.1088/0264-9381/28/19/195004}{\emph{Class. Quant.
  Grav.} {\bfseries 28} (2011) 195004},
  [\href{https://arxiv.org/abs/1103.2307}{{\ttfamily 1103.2307}}].

\bibitem{Anastasiou:2018mfk}
G.~Anastasiou, I.~J. Araya, C.~Arias and R.~Olea, \emph{{Einstein-AdS action,
  renormalized volume/area and holographic R\'enyi entropies}},
  \href{http://dx.doi.org/10.1007/JHEP08(2018)136}{\emph{JHEP} {\bfseries 08}
  (2018) 136}, [\href{https://arxiv.org/abs/1806.10708}{{\ttfamily
  1806.10708}}].

\bibitem{Anastasiou:2020zwc}
G.~Anastasiou, O.~Miskovic, R.~Olea and I.~Papadimitriou, \emph{{Counterterms,
  Kounterterms, and the variational problem in AdS gravity}},
  \href{http://dx.doi.org/10.1007/JHEP08(2020)061}{\emph{JHEP} {\bfseries 08}
  (2020) 061}, [\href{https://arxiv.org/abs/2003.06425}{{\ttfamily
  2003.06425}}].

\bibitem{Maldacena:2011mk}
J.~Maldacena, \emph{{Einstein Gravity from Conformal Gravity}},
  \href{https://arxiv.org/abs/1105.5632}{{\ttfamily 1105.5632}}.

\bibitem{Grumiller:2013mxa}
D.~Grumiller, M.~Irakleidou, I.~Lovrekovic and R.~McNees, \emph{{Conformal
  gravity holography in four dimensions}},
  \href{http://dx.doi.org/10.1103/PhysRevLett.112.111102}{\emph{Phys. Rev.
  Lett.} {\bfseries 112} (2014) 111102},
  [\href{https://arxiv.org/abs/1310.0819}{{\ttfamily 1310.0819}}].

\bibitem{Nester:1991yd}
J.~M. Nester, \emph{{A covariant Hamiltonian for gravity theories}},
  \href{http://dx.doi.org/10.1142/S0217732391003092}{\emph{Mod. Phys. Lett. A}
  {\bfseries 6} (1991) 2655--2661}.

\bibitem{Chen:1998aw}
C.-M. Chen and J.~M. Nester, \emph{{Quasilocal quantities for GR and other
  gravity theories}},
  \href{http://dx.doi.org/10.1088/0264-9381/16/4/018}{\emph{Class. Quant.
  Grav.} {\bfseries 16} (1999) 1279--1304},
  [\href{https://arxiv.org/abs/gr-qc/9809020}{{\ttfamily gr-qc/9809020}}].

\bibitem{Padmanabhan:2011ex}
T.~Padmanabhan, \emph{{Some aspects of field equations in generalised theories
  of gravity}}, \href{http://dx.doi.org/10.1103/PhysRevD.84.124041}{\emph{Phys.
  Rev. D} {\bfseries 84} (2011) 124041},
  [\href{https://arxiv.org/abs/1109.3846}{{\ttfamily 1109.3846}}].

\bibitem{Araneda:2016iiy}
R.~Araneda, R.~Aros, O.~Miskovic and R.~Olea, \emph{{Magnetic Mass in 4D AdS
  Gravity}}, \href{http://dx.doi.org/10.1103/PhysRevD.93.084022}{\emph{Phys.
  Rev. D} {\bfseries 93} (2016) 084022},
  [\href{https://arxiv.org/abs/1602.07975}{{\ttfamily 1602.07975}}].

\bibitem{Ciambelli:2020qny}
L.~Ciambelli, C.~Corral, J.~Figueroa, G.~Giribet and R.~Olea,
  \emph{{Topological Terms and the Misner String Entropy}},
  \href{http://dx.doi.org/10.1103/PhysRevD.103.024052}{\emph{Phys. Rev. D}
  {\bfseries 103} (2021) 024052},
  [\href{https://arxiv.org/abs/2011.11044}{{\ttfamily 2011.11044}}].

\bibitem{Anastasiou:2017mag}
G.~Anastasiou and R.~Olea, \emph{{Holographic correlation functions in Critical
  Gravity}}, \href{http://dx.doi.org/10.1007/JHEP11(2017)019}{\emph{JHEP}
  {\bfseries 11} (2017) 019},
  [\href{https://arxiv.org/abs/1709.01174}{{\ttfamily 1709.01174}}].

\end{thebibliography}\endgroup
\bibliographystyle{JHEP}

\end{document}